\documentclass[prl, reprint, superscriptaddress, showpacs, bibliography]{revtex4-1}
\usepackage[dvips]{graphicx}
\usepackage{color}
\usepackage{amsmath}
\usepackage{amssymb}
\usepackage[breaklinks]{hyperref}
\usepackage[all]{hypcap}

\begin{document}

\title{Direct determination of the Tomonaga-Luttinger parameter $K$ in quasi-one-dimensional spin systems}

\author{Mladen Horvati\'{c}}
\email{mladen.horvatic@lncmi.cnrs.fr}
\affiliation{Laboratoire National des Champs Magn\'etiques Intenses, LNCMI-CNRS (UPR3228),
EMFL, Universit\'e \\ Grenoble Alpes, UPS and INSA Toulouse, Bo\^{i}te Postale 166, 38042 Grenoble Cedex 9, France}
\author{Martin Klanj\v{s}ek}
\affiliation{Jo\v{z}ef Stefan Institute, Jamova c. 39, SI-1000 Ljubljana, Slovenia}
\author{Edmond Orignac}
\affiliation{Universit\'{e} de Lyon, \'{E}cole Normale Sup\'{e}rieure de Lyon, Universit\'{e} Claude Bernard, CNRS, Laboratoire de Physique, 69342 Lyon, France}

\begin{abstract}
We test the analytical formula for the enhancement of the nuclear magnetic resonance rate $T_1^{-1}$ by the critical spin fluctuations, over the simple power-law dependence predicted for a purely one-dimensional spin system, recently derived in the random phase approximation [M. Dupont \textit{et al.}, \href{https://doi.org/10.1103/PhysRevB.98.094403}{Phys. Rev. B {\bf 98}, 094403 (2018)}]. This prediction is experimentally confirmed by excellent fits to the published temperature dependence of $T_1^{-1}$ data in the two representative spin compounds, (C$_7$H$_{10}$N)$_2$CuBr$_4$ (DIMPY) and BaCo$_2$V$_2$O$_8$, providing at the same time a \emph{direct} and convenient experimental determination of the Tomonaga-Luttinger-liquid parameter $K$, very well in agreement with theoretical predictions.
\end{abstract}

\date{\today}

\maketitle

The Tomonaga-Luttinger liquid (TLL) theory provides a general low-energy description, including interactions, for any gapless one-dimensional (1D) system \cite{TheBook}. Its importance in the description of quasi-1D materials is therefore crucial, and it can be regarded as analogous to what the Fermi-liquid description is for three-dimensional (3D) systems. While the main hallmark of the TLL description, namely the power-law dependence of 1D response/correlation functions, had been experimentally well established previously \cite{Schwartz1998, Ishii2003}, it is only a decade ago that quasi-1D quantum spin compounds have provided the final quantitative verification of the TLL theory \cite{Klanjsek2008, Bouillot2011}. In the spin-ladder compound (C$_5$H$_{12}$N)$_2$CuBr$_4$, also known as BPCB, one could compare the experimental values with the TLL-based predictions for the magnetic field ($B$) dependence of \textit{i)} the phase boundary $T_c(B)$ of the low-temperature ($T$) ordered phase, \textit{ii)} the low-$T$ limit of the order parameter of this phase \cite{Klanjsek2008, Blinder2017}, and \textit{iii)} the nuclear magnetic resonance (NMR) spin-lattice relaxation rate $T_1^{-1}(B)$ in the TLL regime, at $T \gg T_c$ \cite{Klanjsek2008, Jeong2016}.
A successful theoretical description of these data thus confirmed the field-induced variations of the two TLL parameters: a dimensionless interaction parameter $K$ that defines the power-law exponents and the renormalized Fermi velocity $u$.
In these systems, $B$ plays the role of the chemical potential controlling the filling of the (spinless) fermion band in the representation onto which the spin system can be mapped. The interaction between fermions depends on the filling of the band, which is notably reflected in the $K(B)$ dependence.
\begin{figure}[b!]
\begin{center}
\includegraphics[width=1.0\columnwidth,clip]{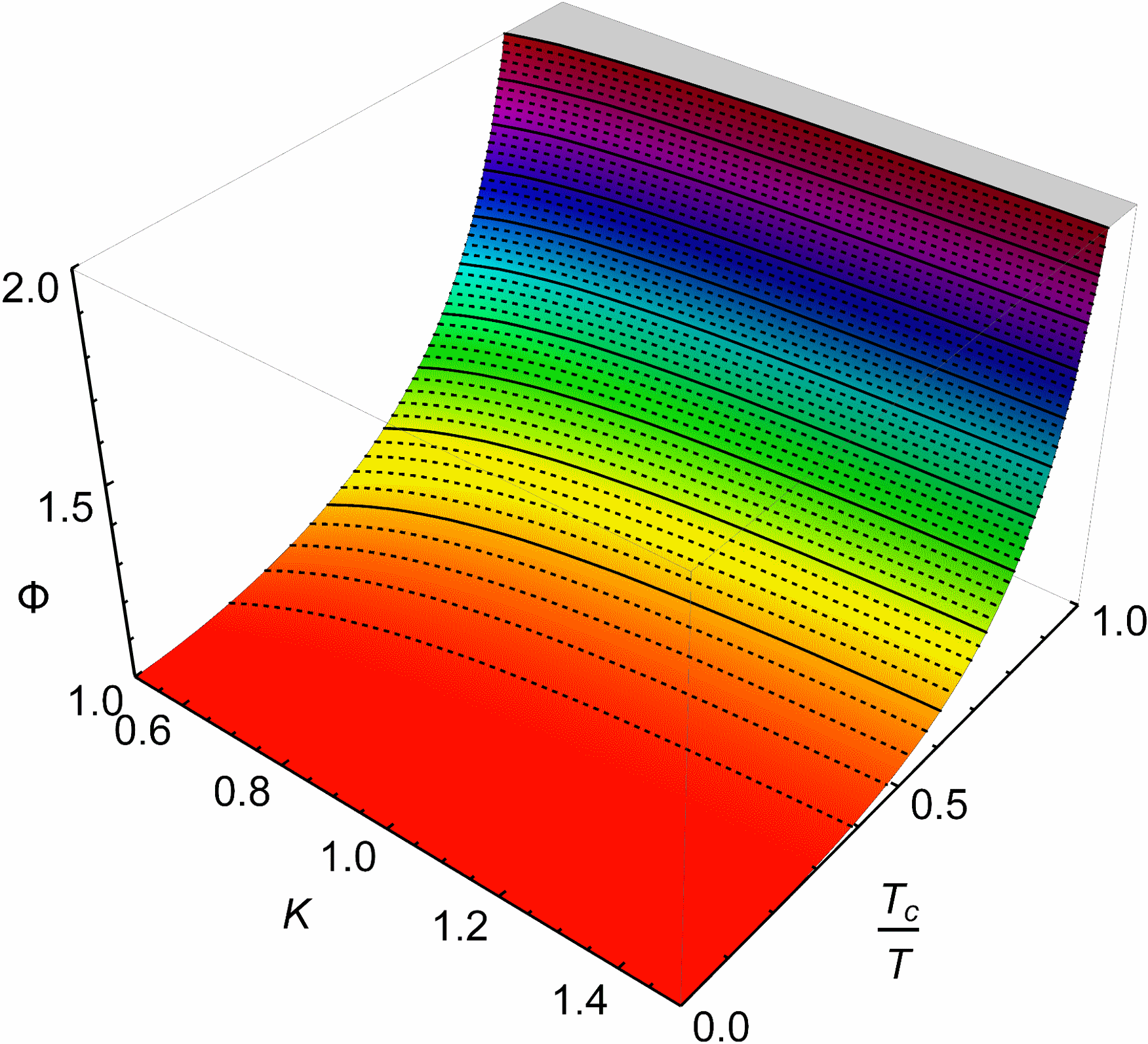}
\caption{The 3D plot of the correction function $\Phi(K,T_c/T)$ defined by Eq.~(\ref{eq2})~\cite{Dupont2018}. The dashed and solid contour lines are spaced at intervals of 0.02 and 0.1, respectively.}
\label{FigPhy}
\end{center}
\end{figure}

However, the first attempt to directly determine the $K$ parameter from the measured $T_1^{-1}(T)$ dependence, performed in the spin-ladder compound (C$_7$H$_{10}$N)$_2$CuBr$_4$, also known as DIMPY, failed \cite{Jeong2013}. This was attributed to the enhancement of relaxation by the critical spin fluctuations in a very broad vicinity of $T_c$. Indeed, a correct determination of the $K$ value from the power-law exponent of the $T_1^{-1}(T)$ temperature dependence is in practice precluded by the enhancement of relaxation related to the nearby $T_c$ on the low-$T$ side, as well as by the inherent limitation of a TLL description to low energy, and thus low temperature, on the high-$T$ side \cite{Coira2016,Dupont2016}. Recently, this was described theoretically both by quantum Monte Carlo (QMC) numerical simulations, and analytically, using the random phase approximation (RPA) to describe the effect of fluctuations \cite{Dupont2018}. The former approach showed that a \emph{purely} 1D (TLL) power-law regime of $T_{1\textrm{TLL}}^{-1}(T) \propto T^{1/2K-1}$ dependence, observed when the three-dimensional (3D) exchange couplings are 3 or more orders of magnitude smaller than 1D coupling, rapidly shrinks and disappears as soon as the 3D couplings strength raises to the level of percent (see Fig.~4 in Ref. \cite{Dupont2018}). In practice, this means that it is not really expected to be observable in most of the experimentally interesting spin systems. Furthermore, a closed analytical expression, depending only on $T_c$ and $K$, was derived within the RPA approximation (and checked against QMC) to take into account the fluctuations related to $T_c$ \cite{Dupont2018}.

Here, we apply this RPA correction to the published $T_1^{-1}(T)$ NMR data in two very different, representative, quasi-1D spin systems \cite{Jeong2013, Klanjsek2015}, and find that it provides a remarkable fit to the data. These fits present the first \emph{direct} experimental determination of the $K$ values that confirms the theoretically predicted values. They also provide a convenient means of the experimental characterization of a quasi-1D system, independent of its complete theoretical description that requires the knowledge of the Hamiltonian and of numerical techniques [QMC, density-matrix renormalization group (DMRG)]. Finally, the fit covers the data quite close to $T_c$ and can also provide an independent estimate of the $T_c$ value. Altogether, it constitutes a reference for the normal quasi-1D behavior, which can be used to reveal nonstandard cases.

In the following, we first discuss the analytical RPA correction to the TLL prediction $T_1^{-1}(T)$, which was cast to a multiplicative correction function $\Phi(K,T_c/T)$ \cite{Dupont2018}:
\begin{widetext}
\begin{eqnarray}
 \label{eq1}
 T_1^{-1}(T,B) &=& T_{1\textrm{TLL}}^{-1}[T,K(B)] \times \Phi[K(B),T_c/T]  =  a\,T^{1/2K-1}\Phi(K,T_c/T)  , \textrm{~where}  \\
\label{eq2}
\Phi(K,T_c/T) &=&
 \frac 1 {N(K)} {\int_{-\infty}^{+\infty}
 \frac{\mathrm{d}\xi}{\sin^2\!\left(\frac {\pi} {8K} \right)+ \sinh^2(\pi \xi)}
 \left|\frac{ \Gamma\!\left( \frac 1 {8K} + i \xi\right)}{\Gamma\!\left(1-\frac 1 {8K} + i \xi\right)} \right|^2
 \frac{\mathrm{E}\!\left[ \left(\frac{T_c}{T}\right)^{4-1/K}
   \left|\frac{\Gamma\left(1-\frac 1 {8K}\right) \Gamma\left(\frac 1 {8K} + i \xi\right)}
     {\Gamma\left(\frac 1 {8K}\right) \Gamma\left(1-\frac 1 {8K} + i \xi\right)} \right|^4 \right]}
 {1 - \left(\frac{T_c}{T}\right)^{4-1/K}
   \left|\frac{\Gamma\left(1-\frac 1 {8K}\right) \Gamma\left(\frac 1 {8K} + i \xi\right)}
     {\Gamma\left(\frac 1 {8K} \right) \Gamma\left(1-\frac 1 {8K} + i \xi\right)} \right|^4}} \\
\textrm{and~} N(K) &=& 2 \Gamma^2\!\left(\frac 1 {4K}\right) \cos\!\left(\frac {\pi} {4K}\right) \mathrm{B}\!\left(\frac 1 {4K}, 1 - \frac 1 {2K}\right).  \nonumber
\end{eqnarray}
\end{widetext}
Here, $\mathrm{E}(x)$ is the complete elliptic integral of the second kind, $\Gamma(x)$ and $\mathrm{B}(x,y)$ are respectively the Euler gamma and beta functions, and $a$ is the amplitude whose magnetic field dependence (not studied here) is determined from the complete expressions for the $T_{1\textrm{TLL}}^{-1}(T,K)$ given in Refs.~\cite{Klanjsek2008, Bouillot2011} (see also Ref. \cite{Hikihara2004}). The correction $\Phi$ depends on $K$ and $T_c$ only, and not on other parameters of the system. Comparison to QMC results showed that the new analytical ``RPA+TLL'' fit is expected to make the experimental determination of $K$ possible even for weakly 1D spin systems, where the ratio of 3D to 1D couplings is as big as 10\% \cite{Dupont2018}.

\begin{figure*}
\begin{center}
\includegraphics[width=1.0\textwidth,clip]{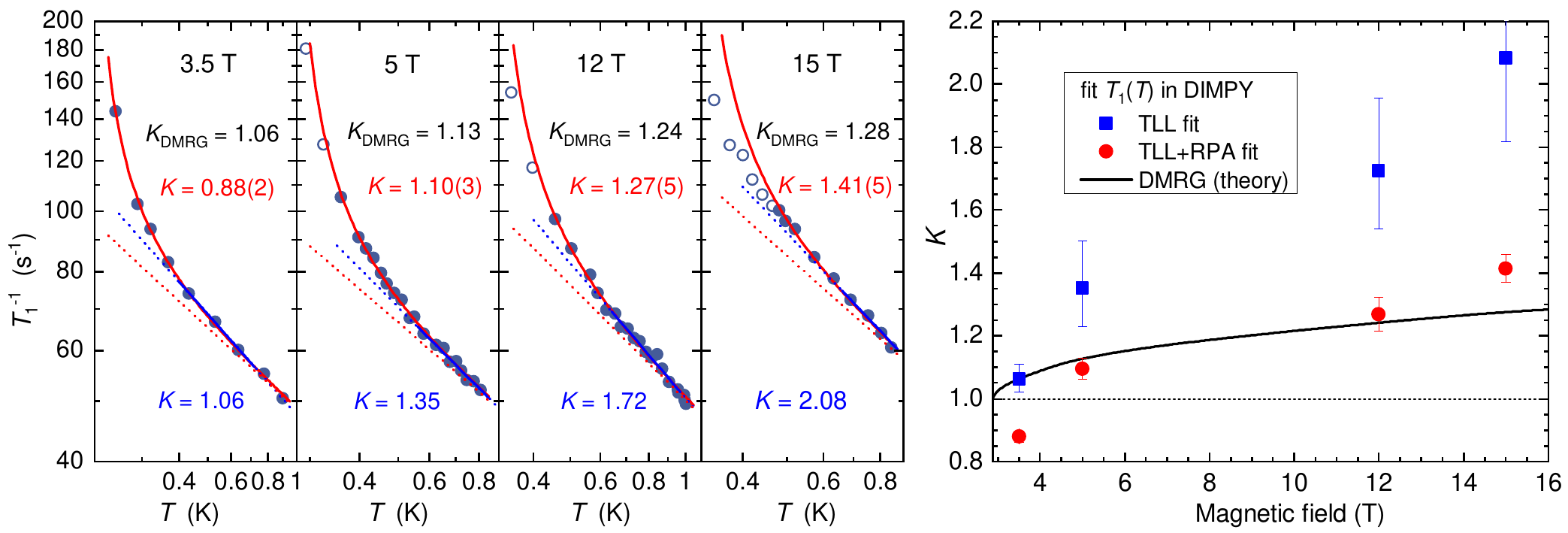}
\caption{(left panels) Comparison of the new RPA+TLL fit (red lines and text) with the previous apparent power-law (TLL) fit (blue lines and text) to the DIMPY data, as given in Ref.~\cite{Jeong2013}. Solid data points denote values taken into account in the former least-squares fit. Pure uncorrected TLL contribution to this fit, $T_{1\textrm{TLL}}^{-1}(T,K) = a\,T^{1/2K-1}$, is given by red dotted lines, to show how much it differs from the apparent power-law fit. The temperature scale of each subpanel starts with the $T_c$ value. $K_{\textrm{DMRG}}$ values refer to the theoretical prediction obtained by DMRG in Ref.~\cite{Jeong2013}, whose field dependence is compared to the experimental $K$ values in the right panel.
}
\label{FigDIMPY}
\end{center}
\end{figure*}

A 3D plot of the $\Phi(K,T_c/T)$ function (Fig.~\ref{FigPhy}) shows that the enhancement of relaxation is moderate, reaching a factor of ~2 at temperature 10\% above $T_c$. Its $K$ dependence is quite feeble, as the contour lines only weakly bend away from the $K$-axis direction. At $T \gtrapprox 2$$T_c$ the enhancement falls below 5\%, and can thus only weakly affect the \emph{field} dependence of $T_1^{-1}$ measured at fixed $T$ well above $T_c$. This \textit{a posteriori} justifies the use of a pure 1D $T_{1\textrm{TLL}}^{-1}$ expression to approximately fit the measured field dependence of relaxation (typically recorded at twice the maximum $T_c$ value) \cite{Klanjsek2008}, also proposed to reveal the attractive ($K > 1$) or repulsive ($K < 1$) nature of a quasi-1D spin system \cite{Jeong2016}.

$\Phi$ is normalized to 1, $\Phi(K,T_c/T \rightarrow 0) \rightarrow 1$, which ensures that the $T_1^{-1}(T)$ on increasing temperature converges to its TLL limit (Fig.~\ref{FigPhy}). However, having a correction of about 5\% at 2$T_c$ means that the apparent power-law fit that neglects the RPA enhancement, taken e.g. in the temperature interval between 2$T_c$ and 3$T_c$, as in the previous analysis of DIMPY data \cite{Jeong2013}, is significantly distorted. For the DIMPY data, this effect is quantified in Fig.~\ref{FigDIMPY}: Indeed, the $K$ values corresponding to the apparent power-law fit are systematically higher than the ones using the RPA+TLL fit defined by Eq.~(\ref{eq2}). For these latter fits, $T_c$ has been determined independently for each field value from the onset of building up of the order parameter, observed through the splitting of the NMR lines (see Fig.~1(b) in Ref.~\cite{Jeong2013}). Only two parameters are then determined by the least-squares fit to the data, the amplitude $a$ and the $K$ value. The $K$ values obtained in this way nicely follow the theoretical prediction, with the exception of the data taken at 3.5~T. We remark that this lowest field value is relatively close to the critical field $B_{c1}$~= 2.9~T, in the vicinity of which the TLL description is not applicable. Finally, the error bars of the RPA+TLL fits are smaller because the temperature interval of these fits is much wider (on the logarithmic scale), which stabilizes the fit.

While the $\Phi(K,T_c/T)$ function (Fig.~\ref{FigPhy}) is weakly dependent on $K$, it clearly diverges as $T$ decreases towards $T_c$ \cite{Giamarchi1999}. Therefore, $T_c$ can be taken as the third free parameter of the fit, in addition to $K$ and $a$ in Eq.~(\ref{eq1}), in order to provide an independent estimate of its value, predicted (extrapolated) from the spin dynamics observed above $T_c$. We present such 3-parameter fits on the example of the published $T_1^{-1}$ data in BaCo$_2$V$_2$O$_8$, an Ising-like $S = 1/2$ spin chain \cite{Klanjsek2015}. For these fits we note that the correction function $\Phi(K,T_c/T)$ has been calculated \cite{Dupont2018} for systems, such as Heisenberg $S = 1/2$ spin ladders, where the dominant spin fluctuations are the antiferromagnetic (AF) transverse ones, which is expected to be valid when $K > 0.5$. It is easy to show that it can also be applied to systems, such as Ising-like chains, where the dominant fluctuations are longitudinal and incommensurate (IC), which is expected to be valid when $K < 0.5$. The formulas that describe the relevant spin correlators and spin susceptibilities for these two types of fluctuations, given by Eqs. (6.47), (6.50) and (6.53) in Ref.~\cite{TheBook}, have an identical form up to the $1/2K \leftrightarrow 2K'$ symmetry transformation/correspondence around a so-called ``$\eta$-inversion'' point at $K = 0.5$ or $\eta = 1$ ($\eta = 1/2K$ \cite{Maeshima2004, Okunishi2007, Klanjsek2015}), at which the dominant fluctuations change their type. As both the RPA correction function and $T_c$ are calculated/defined from the dynamic susceptibility, the same symmetry transformation applies to $\Phi(K,T_c/T)$. Therefore, for the longitudinal IC fluctuations we get:
\begin{equation}
\label{IC}
T_1^{-1}(T) \propto T^{2K-1}\Phi(1/4K,T_c/T) .
\end{equation}

\begin{figure}
\begin{center}
\includegraphics[width=1.0\columnwidth,clip]{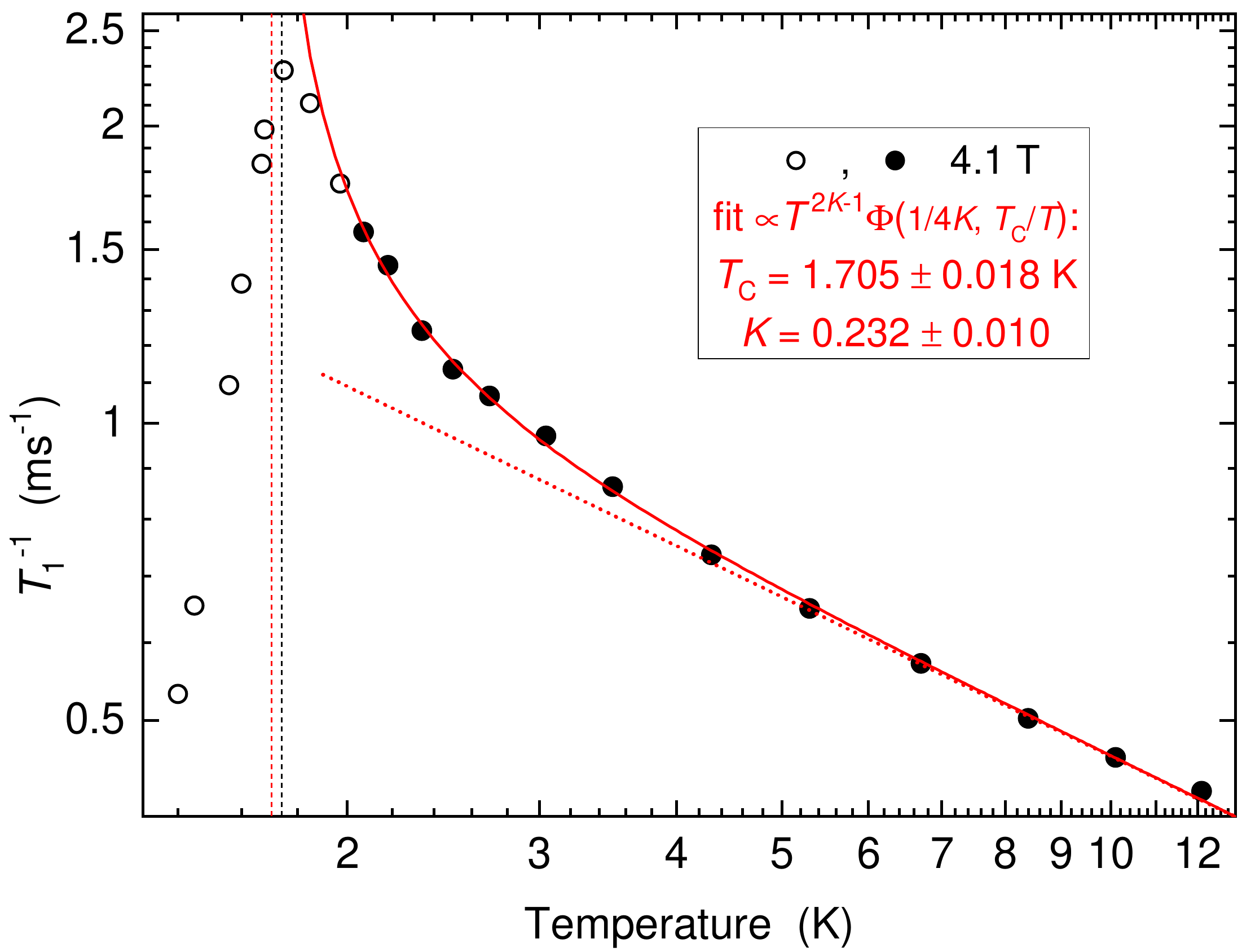}
\caption{The RPA+TLL fit (red solid line) of the BaCo$_2$V$_2$O$_8$ $T_1^{-1}(T)$ data (circles) recorded at 4.1~T \cite{Klanjsek2015}, where solid circles denote the fitted points. The red dotted line is the pure $T_{1\textrm{TLL}}^{-1}$ contribution to this fit. Vertical dashed lines denote the $T_c$ determined by this fit (in red) and from the position of the $T_1^{-1}(T)$ maximum (in black), determined using the spline interpolation through the data points (not shown).}
\label{FigBaCoVOlow}
\end{center}
\end{figure}

Figure \ref{FigBaCoVOlow} shows this fit applied to the BaCo$_2$V$_2$O$_8$ data taken at 4.1~T \cite{Klanjsek2015}, where both $K$ and $T_c$ (and the amplitude) are taken as the fit parameters. Here, the fitted data cover a broad enough temperature interval to well represent both the power-law and the fluctuations-enhanced regime. This is followed by a sharp peak of $T_1^{-1}(T)$, whose maximum reflects the maximum of the critical spin fluctuations and thus precisely defines the $T_c$ value. The corresponding $T_c$ value determined from the RPA+TLL fit given by Eq.~(\ref{IC}) is only 2\% lower, which is within the statistical error as defined by the fit. The equality of these two very different estimates of $T_c$, one reflecting critical dynamics \emph{at} $T_c$ and the other \emph{above} $T_c$, constitutes a very strong confirmation for the validity of the employed correction function. Furthermore, the obtained $K$~= 0.23(1) value is very close to the $K = 1/4$ value expected for the nearby critical field value $B_c$~= 3.8~T. Parenthetically, we observe that the $1/2K \leftrightarrow 2K'$ symmetry connects this value to the noninteracting limit $K'$~= 1.

\begin{figure}
\begin{center}
\includegraphics[width=1.0\columnwidth,clip]{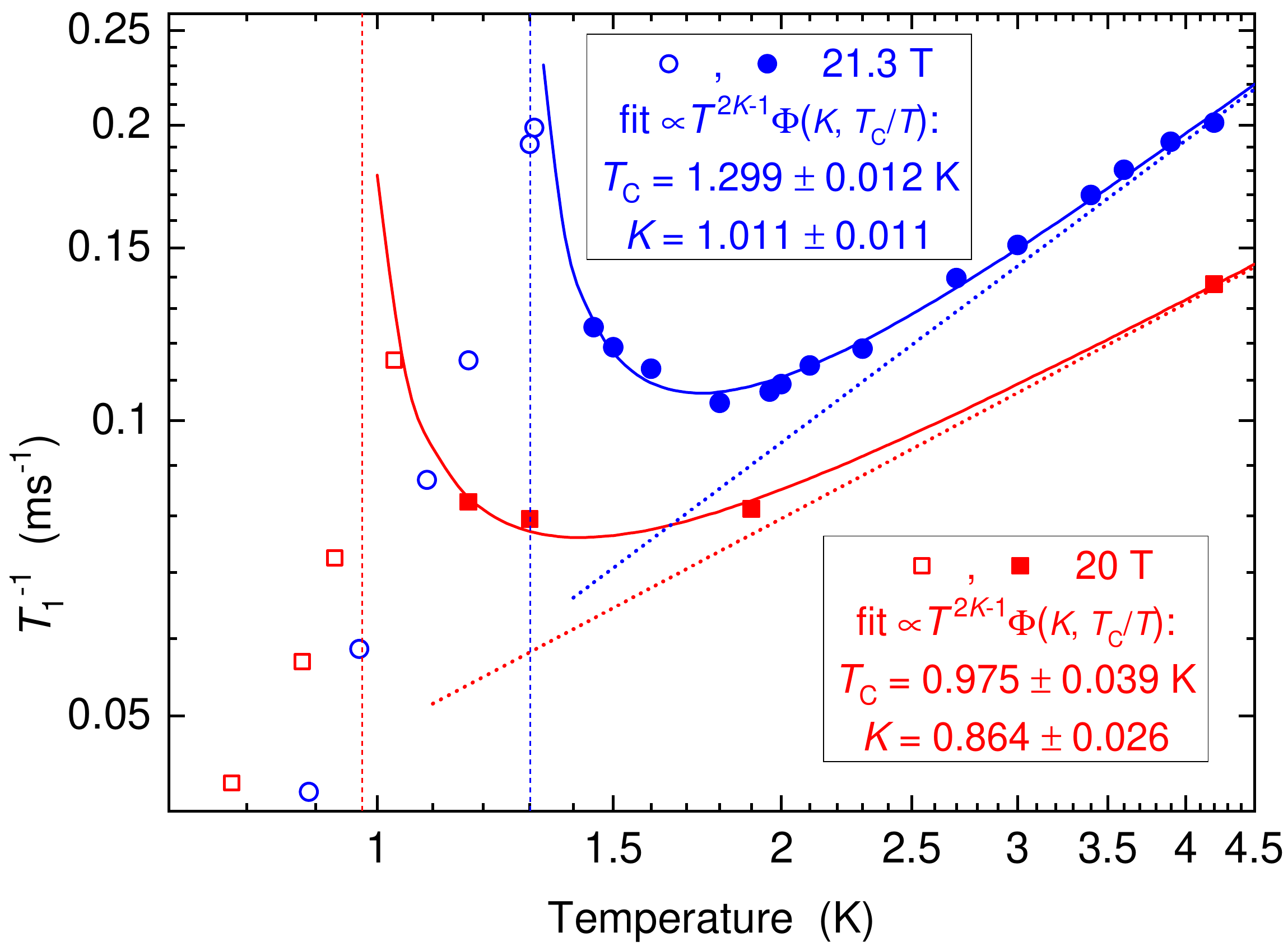}
\caption{The RPA+TLL fits (solid lines) of the BaCo$_2$V$_2$O$_8$ $T_1^{-1}(T)$ data \cite{Klanjsek2015} recorded at 21.3 T (blue color code) and 20~T (red color code), where solid symbols denote the fitted points. Dotted lines are the pure $T_{1\textrm{TLL}}^{-1}$ contribution to this fit. Vertical dashed lines denote the $T_c$ as determined by the fits.}
\label{FigBaCoVOhigh}
\end{center}
\end{figure}

Figure \ref{FigBaCoVOhigh} presents the fits to the two available BaCo$_2$V$_2$O$_8$ data sets close to the saturation field $B_s$~= 22.8~T \cite{Klanjsek2015}. Here, the successful fit is of ``mixed'' character, $T_1^{-1}(T) \propto T^{2K-1}\Phi(K,T_c/T)$: The power-law (TLL) contribution is of the same type as at low fields (Fig.~\ref{FigBaCoVOlow}), corresponding to IC fluctuations, while the correction factor $\Phi$ corresponds to the dominant transverse AF fluctuations, as expected in the $\eta$-inversion scenario \cite{Okunishi2007}. The power-law exponent can then be explained by the nature of the hyperfine coupling in this compound, which filters out the contribution of AF spin fluctuations and thus selects the IC contribution to $T_1^{-1}$ \cite{Klanjsek2015}, even when it is subdominant. As expected, the obtained $K$ values are close to the $K$ = 1 value that is predicted for the nearby saturation field, and they decrease with the field. Therefore, this fit supports the $\eta$-inversion scenario at high fields, as also suggested from the neutron data \cite{Grenier2015}, but unlike the previously proposed interpretation of the NMR data, based on the pure TLL description \cite{Klanjsek2015}. However, for the new fit, it is not clear why the RPA correction factor of the subdominant fluctuations should be the same as for the dominant ones.

The above examples show that the RPA+TLL fit can be successfully applied to cover different types of fluctuations spanning the complete theoretical range of $K$ values in various quasi-1D spin systems. The fit provides the $K$ value that experimentally characterizes a quasi-1D spin system independently of the availability of a theoretical description. The latter can be unavailable because the microscopic Hamiltonian is only partially defined/known, e.g., when the phase diagram extends up to very high magnetic field values that are beyond the current experimental reach. The temperature range successfully covered by the fit typically goes quite close to $T_c$, down to about 1.2$T_c$. This strong extension of the applicable range makes the fit more stable and possible even for systems farther away from the 1D limit. Below $\approx$1.2$T_c$, we expect that the nature of the critical fluctuations changes from the 1D-based one, taken into account by $\Phi$, to the usual 3D fluctuations, whose typical extension in temperature is of the order of 10\%. Additionally, real compounds often present some sort of disorder, leading to a distribution of $T_c$ values and the corresponding broadening of the peak of the measured $T_1^{-1}(T)$ data that reflect the critical fluctuations.

We remark that the $\Phi(K,T_c/T)$ function in principle depends on the geometry of 3D couplings, and that its analytical expression given by Eq.~(\ref{eq2}) has been calculated for the system of tetragonal symmetry \cite{Dupont2018}, see Supplemental Material (SM) for further details \cite{SM}. We have also tested how its form is modified as a function of growing orthorhombic asymmetry (Fig.~\ref{Anisotropy} in \cite{SM}). It turns out that this modification can be neglected up to approximately $J_x/J_y \simeq 2$, a point at which the asymmetry-induced enhancement of the function can be compensated by the effective/fictive increase of the fitted $T_c$ by only 2.6\%. In general, when both the geometry and size of the 3D couplings are known, and their $q_z$ dependence is not frustrated, we can easily compute the \emph{exact} corresponding $\Phi$ function \cite{SM}. However, for most of the real compounds the size of the 3D couplings is \emph{not} known, and we can thus use Eq.~(\ref{eq2}) as a suitable proxy for systems that are \emph{not} strongly anisotropic, and its generalization to the orthorhombic symmetry given by Eq.~(\ref{PhiOrtho}) in SM \cite{SM} to describe other systems. Finally, in SM we also discuss how the $\Phi(K,T_c/T)$ function is evaluated and used in nonlinear fits to $T_1^{-1}(T)$ data in practice, and provide a simple example of the \textit{Wolfram Mathematica} code we used in our fits \cite{SM}.

In conclusion, we performed the first direct comparison between experimentally determined and theoretically predicted values of the parameter $K$ that characterizes the power-law dependences predicted by the TLL description of quasi-1D systems. Using the recently proposed RPA-based correction factor that accounts for the enhancement of the NMR $T_1^{-1}$ rate induced by critical fluctuations \cite{Dupont2018}, we successfully fitted the observed $T_1^{-1}(T)$ dependence in two quasi-1D spin systems, DIMPY and BaCo$_2$V$_2$O$_8$, covering very different regimes of $K$ values. This analysis establishes a simple reference procedure for the characterization of quasi-1D systems. It thus enables us to recognize such systems in compounds whose effective dimension is not evident/known. In particular, it provides a basis to distinguish between quasi-1D and quasi-2D spin systems, whose spin dynamics remains to be characterized. Finally, the RPA correction has been discussed here for the $T_1^{-1}$ data, but it is expected to be relevant to other observables, such as, e.g., specific heat, for which its effect/size remains to be investigated.

\begin{acknowledgments}
We acknowledge valuable discussions with Maxime Dupont, Nicolas Laflorencie, and Mihael Grbi\'{c}.
\end{acknowledgments}

\clearpage

\section{Supplemental material}
\vspace{-0.4cm}
\subsection{to ``Direct determination of the Tomonaga-Luttinger parameter $K$ in quasi-one-dimensional spin systems'' by M. Horvati\'{c}, M. Klanj\v{s}ek, and E. Orignac}

\vspace{0.4cm}
\subsection{$T_1^{-1}(T)$ in the RPA approximation for a quasi-1D system}

We calculate the nuclear spin lattice relaxation rate
\begin{equation}\label{T1def}
    T_1^{-1}\propto T \int{\rm d}^3q\,\,\lim_{\omega \rightarrow 0} \textrm{Im}\,\chi(\textrm{\textbf{q}},\omega)/\omega\,,
\end{equation}
for the dynamic susceptibility given in the random phase approximation (RPA)
\begin{equation}\label{ChiRPA}
    \chi_{\textrm{RPA}}(\textrm{\textbf{q}},\omega) = \frac{\chi_{\textrm{1D}}(q_z,\omega)}{1 + J_{\perp}({\textrm{\textbf{q}}})\,\chi_{\textrm{1D}}(q_z,\omega)}\,,
\end{equation}
where $J_{\perp}({\textrm{\textbf{q}}})$ is the Fourier transform of the transverse (3D) couplings between the 1D systems. As $\hbar\omega_{\textrm{NMR}} \ll k_B T$, where $\omega_{\textrm{NMR}}$ is the NMR resonance frequency, NMR probes the low-energy limit $\omega$\,$\rightarrow$\,0 of the dynamical susceptibility, where $\textrm{Im}\,\chi(\textrm{\textbf{q}},\omega) \propto \omega$. Eq.~(\ref{T1def}) thus reads
\begin{equation}\label{T1RPA}
    T_{1\textrm{RPA}}^{-1} \propto T \int{\rm d}^3q\,\,\frac{\lim_{\omega \rightarrow 0} \textrm{Im}\,\chi_{\textrm{1D}}(q_z,\omega)/\omega}
    {[1 + J_{\perp}({\textrm{\textbf{q}}})\,\chi_{\textrm{1D}}(q_z,0)]^2}\,.
\end{equation}
For a quasi-1D system having dominant transverse ($xx$) staggered (around the antiferromagnetic wave vector $\textrm{\textbf{Q}}$) correlations, the temperature and $K$ dependence of $\chi_{\textrm{1D}}(q_z,\omega)$ are given by Eq.~(6.52) of Ref.~\cite{TheBook}
\begin{flalign}
& \;\;\;\;\;\;\;\;\;\;\;\;\;\;\;\;\chi_{\textrm{1D}}(q_z,\omega) \propto T^{\frac{1}{2K}-1} \, \mathfrak{B}(K,s,w)\,, & \label{Chi1}\\
& \textrm{where}\;\; \mathfrak{B}(K,s,w) = & \label{Chi2}\\
& \;\;\textrm{B}[{\scriptstyle \frac{1}{8K}} + i(s - w), 1 - {\scriptstyle \frac{1}{4K}}]\,\textrm{B}[{\scriptstyle \frac{1}{8K}} + i(s + w), 1 - {\scriptstyle \frac{1}{4K}}]\,, & \nonumber
\end{flalign}

\begin{equation} \label{Chi3}
\;\;s = \hbar u (q_z - Q_z)/(4 \pi k_B T) \,,\;\;w = \hbar \omega/(4 \pi k_B T)\,,
\end{equation}
and $\textrm{B}$ is the Euler Beta function. Inserting Eqs.~(\ref{Chi1})-(\ref{Chi3}) into Eq.~(\ref{T1RPA}) we find for the enhancement of the relaxation due to the presence of $J_{\perp}(\textrm{\textbf{q}})$, that is the ratio $T_{1\textrm{RPA}}^{-1}/T_{1\textrm{TLL}}^{-1} = \Phi$,
\begin{equation}\label{EqPhi}
\Phi(K,\frac{T_c}{T}) = \frac{\int{\rm d}^3q\,\,\frac{\lim_{\omega \rightarrow 0}\textrm{Im}\,\mathfrak{B}(K,q_z,\omega)/\omega}{[1 - (\frac{T_c}{T})^{2-{\scriptscriptstyle \frac{1}{2K}}}\frac{J_{\perp}({\textrm{\textbf{q}}_{\perp},q_z=Q_z)}}{J_{\perp}({\textrm{\textbf{Q}})}}\,\frac{\mathfrak{B}(K,q_z,0)}{\mathfrak{B}(K,0,0)}]^2}}{{\scriptstyle \int{\rm d}q_z\,\,\lim_{\omega \rightarrow 0}\textrm{Im}\,\mathfrak{B}(K,q_z,\omega)/\omega \; \int{\rm d}^2q_{\perp}}}\,.
\end{equation}
We have written the denominator of the first integral using ratios relative to the divergence of $\chi_{\textrm{RPA}}(\textrm{\textbf{Q}},0)$ at $T_c$, namely the point at which 1\,+\,$J_{\perp}({\textrm{\textbf{Q}}})\,\chi_{\textrm{1D}}({\textrm{\textbf{Q}}},0) = 0$. As the $\mathfrak{B}$ function defined by Eq.~(\ref{Chi2}) is strongly localized around $s = 0$ and all the quantities in Eq.~(\ref{EqPhi}) appear as ratios, we can remove $u/T$ from the integral over $q_z$, and the integration over $\textrm{\textbf{q}}$ can thus be taken over the scaled variables without units. That is, Eq.~(\ref{EqPhi}) is precise if $J_{\perp}({\textrm{\textbf{q}}})$ does \emph{not} depend on $q_z$, and it is either a good or a bad approximation when the $q_z$-dependence of $J_{\perp}({\textrm{\textbf{q}}})$ is respectively unfrustrated or frustrated.

For the system of orthorhombic symmetry and the couplings to the four nearest neighbors we have
\begin{equation}\label{JOrtho}
\frac{J_{\perp}({\textrm{\textbf{q}})}}{J_{\perp}({\textrm{\textbf{Q}})}} = \frac{\cos(q_x) + \alpha \,\cos(q_y)}{1 + \alpha} \,,
\end{equation}
where $\alpha = J_y/J_x$ measures the anisotropy, and all the integrals of Eq.~(\ref{EqPhi}) can be taken over the $[0,\pi]$ interval. Integrating over the $q_x$ and $q_y$ variables we get the formula for the enhancement factor in case of orthorhombic symmetry
\begin{widetext}
\begin{flalign}
 & \Phi(K,\frac{T_c}{T}) =
 \frac 1 {N(K)}    \int \frac{d\xi}{\sin^2\!\left(\frac{\pi}{8K} \right)+ \sinh^2(\pi \xi)}\left|\frac{ \Gamma\!\left(\frac{1}{8K} + i \xi\right)}{\Gamma\!\left(1-\frac{1}{8K} + i \xi\right)} \right|^2
  \frac{\mathrm{E}\!\left[\frac{\frac{4\alpha}{(1+\alpha)^2}\left(\frac{T_c}{T}\right)^{4-\frac{1}{K}}h(K,\xi)}{1-\left(\frac{1-\alpha}{1+\alpha}\right)^2\left(\frac{T_c}{T}\right)^{4-\frac{1}{K}}h(K,\xi)}\right]}{\Big[1-\left(\frac{T_c}{T}\right)^{4-\frac{1}{K}} h(K,\xi)\Big] \sqrt{1-\left(\frac{1-\alpha}{1+\alpha}\right)^2\left(\frac{T_c}{T}\right)^{4-\frac{1}{K}}
   h(K,\xi) }} & \nonumber \\
 & \textrm{where}\;\;h(K,\xi)=\left|\frac{\Gamma\!\left(1-\frac{1}{8K}\right) \Gamma\!\left(\frac{1}{8K} + i
          \xi\right)}{\Gamma\!\left(\frac{1}{8K}\right) \Gamma\!\left(1-\frac{1}{8K} + i \xi\right) } \right|^4\;\;\textrm{and}\;\;N(K) = 2 \Gamma^2\!\left(\frac 1 {4K}\right) \cos\!\left(\frac {\pi} {4K}\right) \mathrm{B}\!\left(\frac 1 {4K}, 1 - \frac 1 {2K}\right)\,. & \label{PhiOrtho}
\end{flalign}
\end{widetext}
For the tetragonal symmetry, where $\alpha = 1$, this expression obviously reduces to the one given by Eq.~(\ref{eq2}) of the main manuscript.

While in principle a numerical evaluation of a 1D integral given by Eq.~(\ref{PhiOrtho}) should be faster than the evaluation of the ``original'' 3D integral given by the Eq.~(\ref{EqPhi}), in practice it turns out that, e.g., \textit{Wolfram Mathematica} software handles the latter integral fast enough for normal usage. Direct usage of the Eq.~(\ref{EqPhi}) is thus preferred, as the corresponding code is simpler to write, and we can as well implement other geometries of the $J_{\perp}({\textrm{\textbf{q}}})$ couplings.

Finally, when the nuclear site that we use for recording the $T_1^{-1}$ data is coupled to several different electronic spins, its hyperfine coupling becomes ${\textrm{\textbf{q}}}$-dependent, which introduces the filtering factor $\mathfrak{F}({\textrm{\textbf{q}}})$ in the integral over ${\textrm{\textbf{q}}}$
\begin{equation*}
    T_1^{-1}\propto T \int{\rm d}^3q\,\,\mathfrak{F}({\textrm{\textbf{q}}})\,\lim_{\omega \rightarrow 0} \textrm{Im}\,\chi(\textrm{\textbf{q}},\omega)/\omega\,.
\end{equation*}
In the numerical evaluation of Eq.~(\ref{EqPhi}), the corresponding modification can be taken into account, and is expected to be important if $\mathfrak{F}({\textrm{\textbf{Q}}})\sim0$.

\begin{figure}[b!]
\begin{center}
\includegraphics[width=1.0\columnwidth,clip]{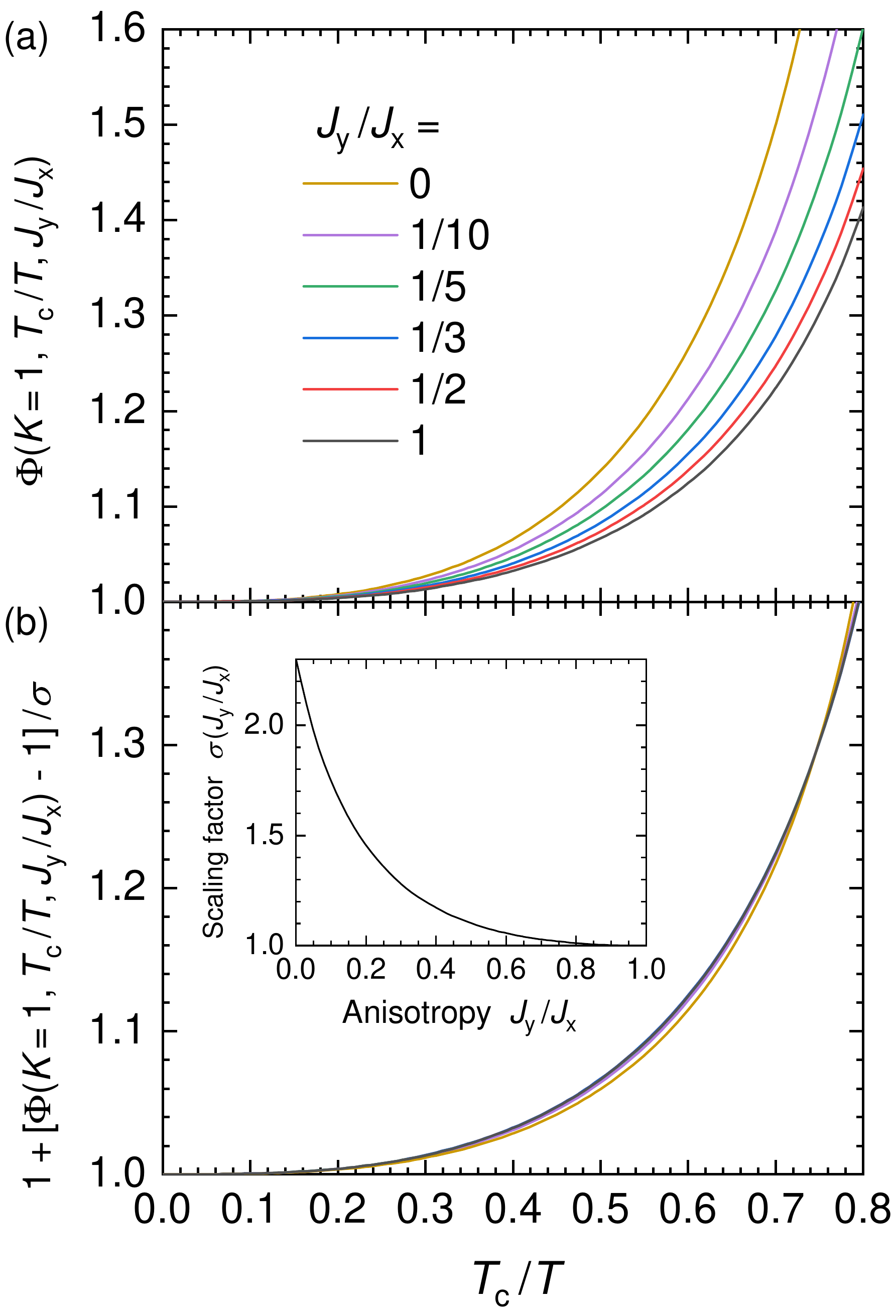}
\caption{(a) The anisotropy dependence of the RPA correction factor and (b) the overlap of these curves by scaling the $(\Phi - 1)$ values by the factor given in the inset (see the text).}
\label{Anisotropy}
\end{center}
\end{figure}

\subsection{Anisotropy dependence of the RPA correction factor for an orthorhombic system}

Fig.~\ref{Anisotropy}(a) shows the anisotropy $\alpha = J_y/J_x$ dependence of the RPA correction factor $\Phi(K$\,=\,$1,T_c/T)$ for the $J_{\perp}({\textrm{\textbf{q}}})$ coupling defined by Eq.~(\ref{JOrtho}). As the $K$ dependence of $\Phi$ is quite weak, the presented results for $K=1$ are very representative for the whole relevant range of $K$ values. We can see that the growing anisotropy enhances $\Phi$. This is indeed expected, as we are in fact approaching the case of 2D, where the fluctuations should be enhanced. Nevertheless, the effect remains quite small up to $\alpha \approx 1/2$. Apparently, the family of curves shown in Fig.~\ref{Anisotropy}(a) can be superposed by scaling their $(\Phi - 1)$ values, and we did that using the least-squares fit in the [0.0, 0.8] interval, see Fig.~\ref{Anisotropy}(b). The overlap is nearly perfect for all the curves, with the exception of the 2D limit, $J_y = 0$. We can thus fit \emph{any} orthorhombic system using simply the $[\sigma(\Phi - 1) + 1]$ scaling based on the $\Phi$ function given by Eq.~(\ref{eq2}) of the main manuscript, and thereby experimentally determine the anisotropy of the system from the value of the fitted scaling parameter $\sigma$. As regards the 2D limit ($J_y = 0$), we might doubt its validity, because the 1D susceptibility $\chi_{\textrm{1D}}(q_z,\omega)$ should no longer be a good starting point of the RPA approximation to describe purely 2D fluctuations. Finally, we can equally well superpose the $\Phi$ curves shown in Fig.~\ref{Anisotropy} by extending their $T_c/T$ scale (not shown). In this way we find that the corresponding effective/fictive modification of $T_c$ amounts to only +2.6\% for $\alpha = 1/2$, which is comparable to the error bars on $T_c$ determination. This means that a weak anisotropy, i.e., the values $1 \geq \alpha \gtrapprox 1/2$, will be practically undetectable by $T_1^{-1}(T)$ data.

\subsection{Numerical evaluation of the RPA correction factor}

Both Eq.~(\ref{PhiOrtho}) and Eq.~(\ref{EqPhi}) can be \emph{literally} converted into a very compact, e.g., \textit{Wolfram Mathematica} code, and the numerical evaluation of the $\Phi$-function values is quite fast. However, when such an integral function is further employed in a nonlinear fit, the latter becomes inconveniently slow. We thus find it better to first tabulate the function values over a convenient domain of variables and then use in fitting the function redefined as the polynomial interpolation over the tabulated values.

As an example of such a fit, in the following page, we provide a copy of the \emph{Mathematica} code for the fit of BaCo$_2$V$_2$O$_8$ data at 4.1~T that is presented in Fig.~\ref{FigBaCoVOlow} of the main manuscript.

\clearpage
~
\includegraphics[width=0.96\textwidth,clip]{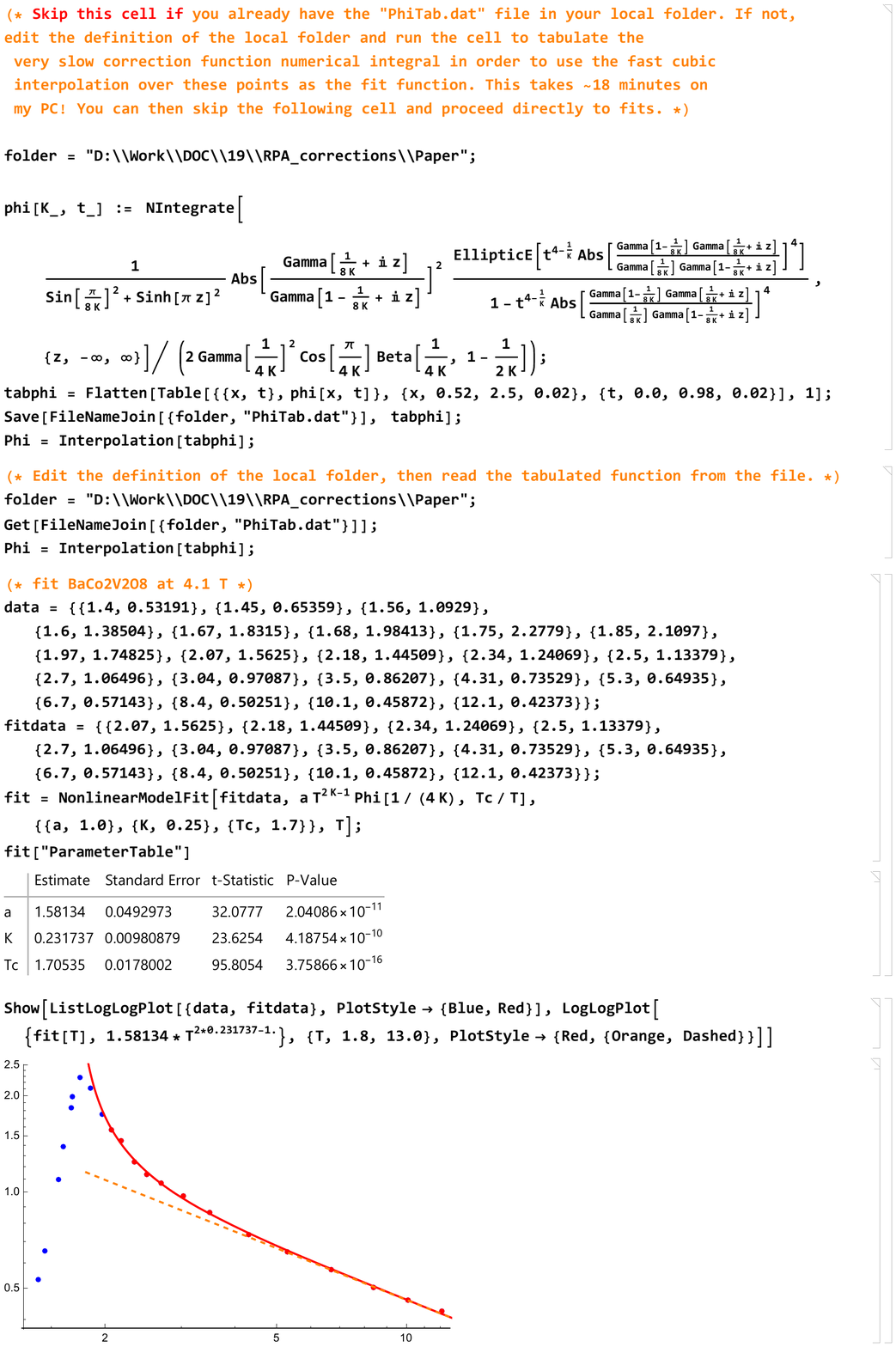}
~
\end{document}